\documentclass[letterpaper,journal]{IEEEtran}
\usepackage{amsmath,amsfonts}
\usepackage{algorithmic}
\usepackage{algorithm}
\usepackage{array}
\usepackage[caption=false,font=normalsize,labelfont=sf,textfont=sf]{subfig}
\usepackage{textcomp}
\usepackage{stfloats}
\usepackage{url}
\usepackage{verbatim}
\usepackage{graphicx}
\usepackage{xcolor}
\usepackage{soul}
\usepackage{cite}
\usepackage{acro}
\DeclareAcronym{RAN}{
    short = RAN ,
    long  = Radio Access Network
}
\DeclareAcronym{UE}{
    short = UE ,
    long  = User Equipment ,
    long-plural-form = User Equipment
}
\DeclareAcronym{UAV}{
    short = UAV ,
    long  = Unmanned Aerial Vehicle
}
\DeclareAcronym{HAP}{
    short = HAP ,
    long  = High Altitude Platform
}
\DeclareAcronym{LAP}{
    short = LAP ,
    long  = Low Altitude Platform
}
\DeclareAcronym{NTN}{
    short = NTN ,
    long  = Non-Terrestrial Networks
}
\DeclareAcronym{PPDR}{
    short = PPDR ,
    long  = Public Protection and Disaster Relief
}
\DeclareAcronym{QoS}{
    short = QoS ,
    long  = Quality of Service
}
\DeclareAcronym{CPU}{
    short = CPU ,
    long  = Central Processing Unit
}
\DeclareAcronym{GPU}{
    short = GPU ,
    long  = Graphics Processing Unit
}
\DeclareAcronym{FPGA}{
    short = FPGA ,
    long  = Field-Programmable Gate Array
}
\DeclareAcronym{TPU}{
    short = TPU ,
    long  = Tensor Processing Unit
}
\DeclareAcronym{SmartNIC}{
    long  = Smart Network Interface Card ,
    short = SmartNIC
}

\DeclareAcronym{SBA}{
    short = SBA ,
    long  = Service-Based Architecture
}

\DeclareAcronym{CDO}{
    short = CDO ,
    long  = Cross-Domain Orchestrator
}

\DeclareAcronym{IDO}{
    short = IDO ,
    long  = Intra-Domain Orchestrator
}

\DeclareAcronym{LLM}{
    short = LLM ,
    long  = Large Language Model
}

\DeclareAcronym{RL}{
    short = RL ,
    long  = Reinforcement Learning
}

\DeclareAcronym{NMS}{
    short = NMS ,
    long  = Network Management System
}

\DeclareAcronym{API}{
    short = API ,
    long  = Application Programming Interface
}

\DeclareAcronym{AI}{
    short = AI ,
    long  = Artificial Intelligence
}

\DeclareAcronym{E2E}{
    short = E2E ,
    long  = end-to-end
}

\DeclareAcronym{SOAR}{
    short = SOAR ,
    long  = {Security Orchestration, Automation, and Response}
}

\DeclareAcronym{ZTP}{
    short = ZTP ,
    long  = zero-touch provisioning
}

\DeclareAcronym{ZSM}{
    short = ZSM ,
    long  = Zero-touch network and Service Management
}

\DeclareAcronym{HPA}{
    short = HPA ,
    long  = Horizontal Pod Autoscaling
}

\DeclareAcronym{A2A}{
    short = A2A ,
    long  = Agent-to-Agent
}

\DeclareAcronym{SLM}{
    short = SLM ,
    long  = Small Language Model
}

\begin{document}

\title{Conversational Orchestration for Organic 6G}

\author{
  \IEEEauthorblockN{
    Masoud Shokrnezhad\textsuperscript{1} and Tarik Taleb\textsuperscript{2}
  }
  \IEEEauthorblockA{
    \\
    \textsuperscript{1} \textit{ICTFICIAL Oy, Espoo, Finland}; masoud.shokrnezhad@ictficial.com \\
    \textsuperscript{2} \textit{Ruhr University Bochum, Bochum, Germany}; tarik.taleb@rub.de \\
    \vspace{-20pt}
  }
}



\maketitle

\begin{abstract}
  The Organic 6G vision of a network of networks spanning an edge--cloud continuum complemented by non-terrestrial resources requires, to realize its promise, service provisioning that is simple to operate, scalable across independently administered domains, and agile under domain churn (i.e., domains dynamically joining and leaving). Despite advances in cross-domain orchestration, many proposals rely on heavy integration fabrics, multi-layer coordinators, and deep telemetry pipelines that hinder deployability and amplify coordination overhead. We propose a lightweight, decentralized conversational orchestration framework based on Large Language Model (LLM)-driven domain agents. Each domain remains autonomous: an agent observes local state via tools, reasons in a closed loop, and exchanges summaries with neighboring agents over an Agent-to-Agent (A2A) overlay aligned with data-plane coupling. Fast feasible placement is enabled by periodic, routing-like dissemination of reachability advertisements (latency, bottleneck bandwidth, and compute capacity), while safe re-optimization, scaling, and migration are handled through event-driven requests and negotiation. To meet real-time constraints, we deploy a compact reasoning model trained with verifier-based self-verification and periodically refined online via shadow updates. Simulations show manageable, near-linear control-plane overhead as domains scale and during domain joins, and robust decision quality, including recovery after objective changes. We close by outlining future research directions for principled, secure, and uncertainty-aware agentic orchestration in Organic 6G.

\end{abstract}


\section{Introduction}
Organic 6G envisions the future mobile system as a software-centric \emph{network of networks}, where heterogeneous radio, compute, and transport resources, spanning edge-to-cloud and including \ac{NTN} assets, continuously appear, disappear, and reshape under independent administrative control \cite{corici_organic_2023}. This evolution aligns with the broader push toward \emph{open networks} built on interoperable standards for cross-domain service delivery \cite{ghosh_future_2025}. From a multi-domain perspective, this direction is not optional: end-to-end services inevitably traverse domains with distinct policies, trust boundaries, and operational constraints. Hence, 6G must be designed to function under decentralization and churn, supporting seamless domain onboarding/offboarding rather than assuming a fixed, single-operator topology. This structural dynamism (an infrastructure that must remain coherent and serviceable as resource domains continuously join, leave, and evolve) is the central challenge motivating this paper: any orchestration approach that assumes a static, globally known topology will fail in this setting.

However, an Organic 6G infrastructure is only useful if services can be provisioned over it. In this paper, \emph{service provisioning} denotes the end-to-end process of turning a service description into running, reachable service instances by (i) selecting and placing compute across the edge--cloud continuum and (ii) establishing the required connectivity so users can be bound to those instances, while continuously adapting via scaling and migration as conditions change. Doing so across many independently administered domains requires methods that are \textbf{scalable} (bounded coordination cost as domains grow), \textbf{simple} (deployable and failure-resilient without heavy operational burden), and \textbf{agile} (plug-and-play domain join/leave, i.e., resource domains dynamically onboarding or going offline at runtime) \cite{noauthor_advanced_nodate}. While recent work has made progress toward cross-domain orchestration in 5G/6G settings, existing approaches often presume substantial architectural machinery (e.g., multi-layer coordinators, integration fabrics, deep telemetry/\ac{AI} pipelines) and pre-established federation agreements, which complicate plug-and-play dynamics and can drive coordination overhead upward with the number of domains \cite{giannopoulos_across_2023, molner_aiora_2025, dalgitsis_cloud-native_2024, benlloch-caballero_e2e_2025, 9247517}.

To fill this gap, this paper proposes a lightweight, decentralized \emph{conversational orchestration} approach based on \ac{LLM}-driven domain agents that coordinate service provisioning decisions over an inter-domain overlay graph.
The key insight is that LLM-based agents shift orchestration complexity from engineered architectural machinery (integration fabrics, coordinator hierarchies, raw telemetry aggregation) to goal-driven reasoning.
Each domain remains autonomous -- its agent optimizes locally using domain tools and policies -- yet end-to-end coordination emerges by exchanging only goal-driven, summarized reachability information with neighboring agents via \ac{A2A} messaging.
Concretely, the proposed design combines (i) periodic, routing-like dissemination of compact resource reachability advertisements and (ii) on-demand request/negotiation for safe re-optimization and migration, thereby keeping coordination bounded while supporting domain churn. This keeps the design \textbf{scalable} (localized exchanges, no centralized entity), \textbf{simple} (no heavy orchestration machinery), and \textbf{agile} (plug-and-play domain join/leave).

The remainder of this paper is organized as follows: Section~II provides background and requirements; Section~III presents the proposed agent-based architecture and decision process; Section~IV evaluates the approach via simulations; and Section~V discusses future directions and concludes this paper.

\section{Background}
\subsection{Infrastructure}
From a resource view, Organic 6G treats the infrastructure as a distributed continuum of connectivity and computation rather than a fixed, centralized stack. At the access edge, the \ac{RAN} (including disaggregated/vRAN and Open RAN components) anchors the system by attaching \acp{UE} and exposing radio resources that can be shaped per service. Beyond the \ac{RAN}, computation spans in-network elements, near-edge nodes (co-located with or adjacent to the \ac{RAN}), far-edge nodes (district/metro-level aggregation points), regional micro-data centers, and central clouds, forming a continuum that trades latency for capacity.
The interconnect is inherently heterogeneous: front-/mid-/backhaul spans fiber, copper, terrestrial wireless, and \ac{NTN} assets such as satellites, \acp{UAV}, and \acp{HAP}/\acp{LAP}, each with distinct bandwidth, latency, and intermittency profiles \cite{shokrnezhad_toward_2025}; this heterogeneity extends reach and resilience where terrestrial links are sparse or disrupted.
Hardware heterogeneity (\acp{FPGA}, \acp{SmartNIC}, GPUs) further supports efficient data-plane execution at the edge.

\begin{figure}[t]
    \centering
    \includegraphics[width=0.9\linewidth]{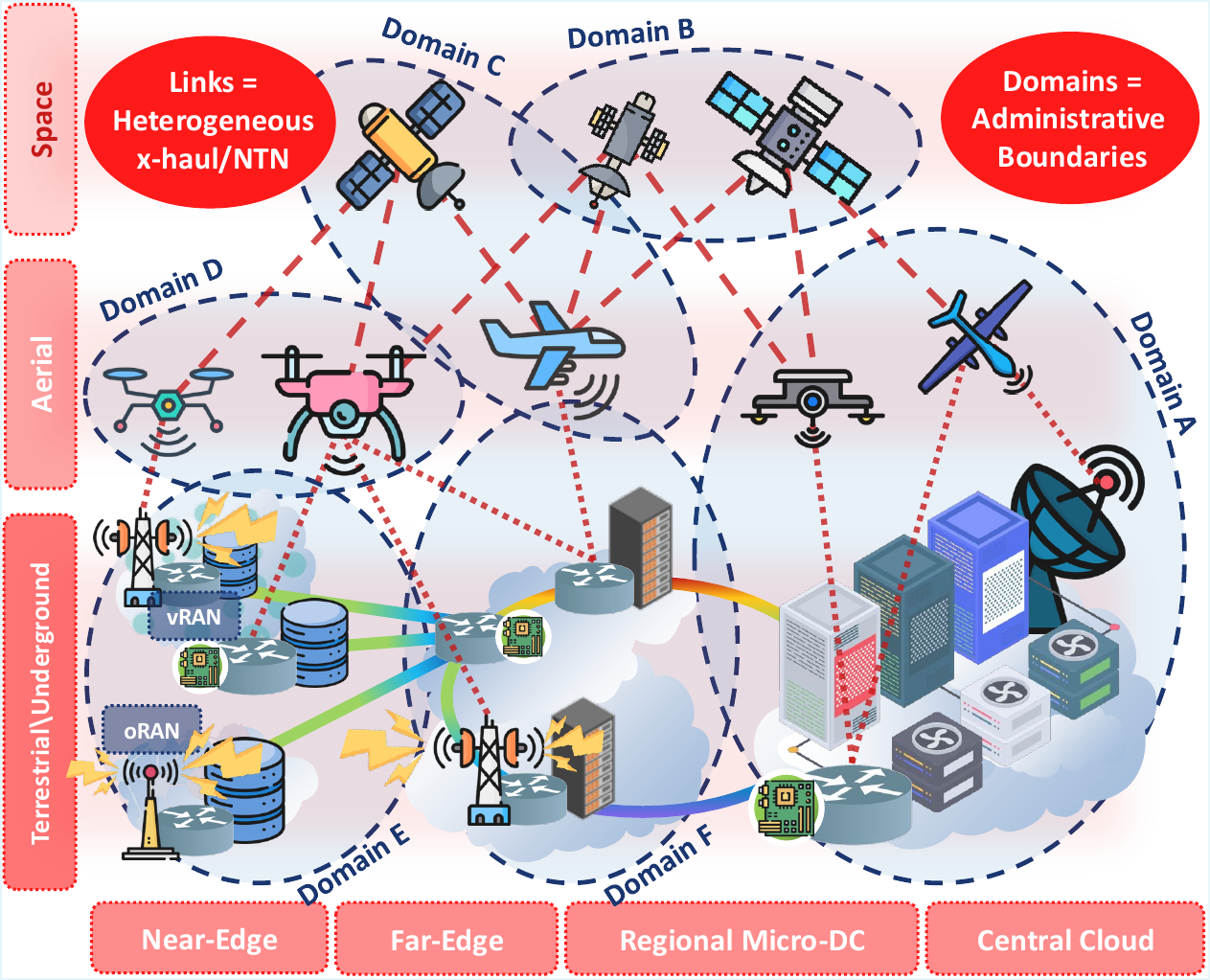}
    \vspace{-8pt}
    \caption{Organic 6G infrastructure as a distributed continuum of connectivity and computation spanning terrestrial, aerial, and space assets from near-edge to central cloud, organized as a modular network of networks with isolated administrative domains interconnected via heterogeneous x-haul/\ac{NTN}.}
    \vspace{-0.3cm}
    \label{fig:organic6g_infrastructure}
\end{figure}

At a system level, Organic 6G interprets this continuum as a modular \emph{network of networks} built from isolated administrative domains, where each domain can own and operate a well-scoped subset of the overall resource pool. In practice, one entity may provide the access segment (e.g., a local \ac{RAN} deployment), another may operate a portion of the compute continuum (e.g., near- or far-edge nodes, or regional micro-data centers), while additional entities may contribute transport/x-haul capacity or non-terrestrial resources. Each domain encapsulates its resources behind domain-local policies and management functions, enabling independent lifecycle management without assuming a static, globally engineered topology. The result is an infrastructure that remains structurally coherent even when its constituent resource domains appear, disappear, or change shape over time. The co-existence of these heterogeneous resources, each operated by an independent administrative domain with its own policies and capabilities, is what makes end-to-end service provisioning non-trivial: a single service may require compute at multiple tiers, connectivity across heterogeneous link types, and continuous adaptation as users move or conditions change, all without any single entity having global visibility or control. Fig.~\ref{fig:organic6g_infrastructure} provides an overview of the above. Unlike ephemeral networks (temporary by design) or end-to-end network slicing, which virtualizes resources over a globally orchestrated substrate, Organic 6G describes a persistent, multi-domain network of networks where the physical infrastructure itself evolves dynamically under independent administrative control, with no single entity holding a global view.

\subsection{Service Provisioning}
Building on the infrastructure continuum described above, service provisioning refers to the end-to-end process of turning a service description into running, reachable service instances over the available \emph{computing} and \emph{networking} resources. For example, a latency-sensitive service may be placed on a near-edge node in the user's current domain, with connectivity established through the local \ac{RAN} and x-haul; as the user moves across domain boundaries, the instance is migrated to a near-edge node in the new domain to preserve \ac{QoS}. Concretely, it entails placing service instances across the edge--cloud spectrum and establishing \ac{RAN}/x-haul connectivity to bind users to those instances \cite{farhoudi_service_2025}. In Organic 6G, provisioning is driven by stringent \ac{QoS} objectives under user mobility and highly dynamic resource conditions: latency is often the dominant metric for interactive and control workloads, while availability captures whether sufficient compute and connectivity resources remain continuously accessible to sustain a user’s service experience.

Beyond initial placement, service provisioning includes lifecycle adaptation operations that keep the service performant as conditions change. Scaling adjusts the number of active instances and their allocated resources to follow demand and to mitigate hotspots, whereas migration relocates an instance (or its execution context) to a different node to preserve \ac{QoS} as users move or as local resources become scarce or unreliable. Together, placement, binding, scaling, and migration form the core of service provisioning in Organic 6G: because the environment is highly dynamic (user loads shift, users move, and infrastructure conditions fluctuate), a static one-time placement is insufficient, and continuous adaptation is required to maintain a valid mapping between users, service instances, and infrastructure resources so that \ac{QoS} objectives remain satisfied. Section~III describes how the proposed agent-based approach addresses these challenges through continuous reachability tracking, event-driven re-optimization, and online model refinement.

\subsection{Problem Statement \& Challenges}
The core problem addressed in this paper is the optimization of service provisioning over an Organic 6G infrastructure. Given the distributed pool of \emph{computing} resources (from near-edge to central and non-terrestrial nodes) and \emph{networking} resources (the \ac{RAN} and heterogeneous transport/x-haul), an orchestrator must (i) collect and maintain sufficiently accurate system information across the network of networks, (ii) decide provisioning actions, including placement, user--instance binding, scaling, and migration, and (iii) enforce these actions on the underlying resources. This closed-loop must operate under stringent \ac{QoS} requirements (e.g., latency) while the system exhibits high dynamism. Importantly, any practical solution must satisfy three system-level requirements beyond single-domain optimality \cite{noauthor_advanced_nodate}: \textbf{scalability} (coordination cost must remain bounded as domains grow, avoiding centralized bottlenecks and single points of failure); \textbf{simplicity} (the architecture must be deployable and failure-resilient without heavy operational burden); and \textbf{agility} (domains must support plug-and-play onboarding/offboarding so the system remains functional as resource domains appear, disappear, or change shape).

Recent work has investigated cross-domain orchestration for service provisioning in multi-domain 5G/6G environments. Giannopoulos \textit{et al.} \cite{giannopoulos_across_2023} proposed ACROSS, a two-level architecture where domain orchestrators are supervised by a cross-domain coordinator through a standardized integration fabric, relying on deep telemetry and \ac{AI}-assisted zero-touch provisioning automation. Molner \textit{et al.} \cite{molner_aiora_2025} introduced AIORA, advocating virtual continuums spanning multiple segments and a cross-segment coordination substrate built atop open interfaces and nested \ac{AI}-driven closed loops. Dalgitsis \textit{et al.} \cite{dalgitsis_cloud-native_2024} addressed inter-operator continuity via cloud-native slice federation, translating slice templates into federated templates exchanged over operator federation interfaces to instantiate slices in visited domains. Benlloch-Caballero \textit{et al.} \cite{benlloch-caballero_e2e_2025} studied end-to-end slicing for security automation, coupling distributed sensing and topology tracking with an orchestration loop and cross-domain enforcement agents. Finally, Santos \textit{et al.} \cite{9247517} proposed a hierarchical orchestration scheme for end-to-end network slicing, introducing a higher-level hyperstrator to coordinate distributed per-segment orchestrators across domains, while relying on pre-defined inter-orchestrator interfaces and a central coordination point for lifecycle management. While the achievements are valuable, they still fall short of the \textbf{simplicity}, \textbf{agility}, and \textbf{scalability} requirements emphasized above: they typically presume substantial architectural machinery (integration fabrics, multiple orchestration layers, deep telemetry/\ac{AI} pipelines, or custom enforcement components), pre-established interoperability and federation agreements that complicate plug-and-play domain join/leave, and coordination patterns whose overhead exponentialy grows with the number of domains and slices (e.g., centralized supervision, heavy telemetry aggregation, or increasing inter-orchestrator messaging).

These gaps motivate our lightweight, decentralized agent-based coordination approach in Section~III, designed to keep coordination bounded, operational complexity low, and domain dynamics seamless at Organic 6G scale.

\section{Proposed Approach}
\subsection{Architecture}
\begin{figure}[t]
    \centering
    \includegraphics[width=0.9\linewidth]{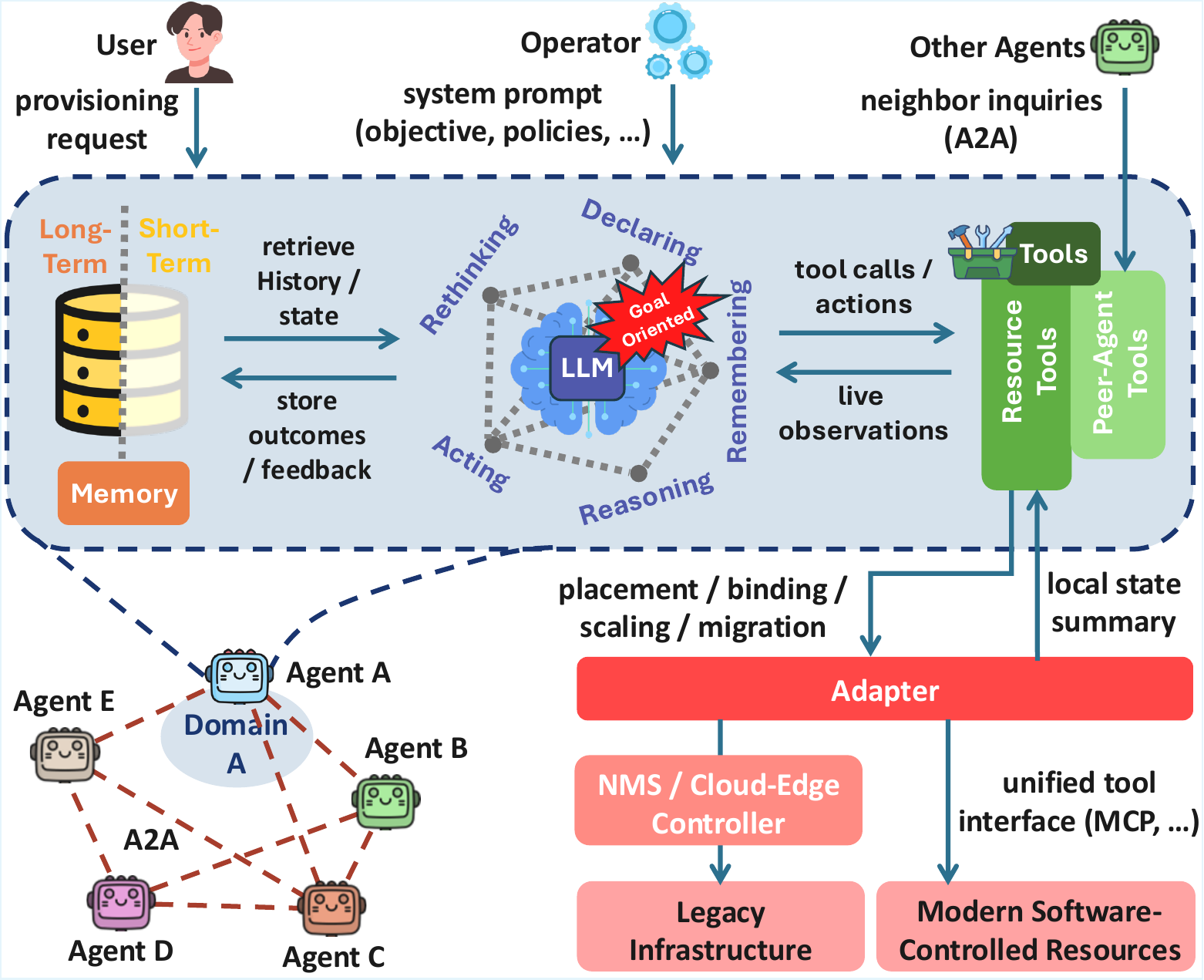}
    \vspace{-8pt}
    \caption{Illustration of the proposed \ac{LLM}-based domain agent and its inter-domain overlay control plane for Organic 6G service provisioning. The \ac{LLM} is the reasoning core; memory provides working context and long-term history; the tool layer interfaces with domain resources and peer agents via \ac{A2A}; and the adapter bridges the agent to both modern software-defined infrastructure (e.g., via MCP) and legacy \ac{NMS}/\ac{API} stacks.}
    \vspace{-0.5cm}
    \label{fig:agentic_architecture}
\end{figure}

We build our solution around \emph{\ac{LLM}-based agents}: autonomous control entities that observe their environment, reason, and act in closed loop. \acp{LLM} are chosen over analytical or pure data-driven alternatives because their goal-driven reasoning generalizes across heterogeneous domain configurations and adapts to objective changes without re-engineering (the accompanying \ac{RL}-based specialization then ensures task-specific optimization competence).
Here, the \ac{LLM} serves as the agent's reasoning core with five internal capabilities: \emph{goal setting} (what to optimize), \emph{decision synthesis} (generating candidate actions), \emph{memory} (storing and retrieving state and feedback), \emph{action execution} (enforcing decisions via tools), and \emph{self-correction} (revising the reasoning path based on outcomes). These capabilities are realized with three modules: the \ac{LLM} itself, a memory store (short-term working context plus persistent long-term history), and a tool layer that interfaces with resources and peer agents. An overview of the proposed agent architecture is shown in Fig~\ref{fig:agentic_architecture}. This design is deliberately goal-oriented: rather than relying on predefined orchestration pipelines, the agent \emph{selects which observations to request}, \emph{which tools to invoke}, and \emph{which control actions to apply} based on the currently declared goals and constraints.
The claimed system properties (simplicity, agility, scalability) are architectural properties of the orchestration system, not of the model: the model's computational cost is local to each domain and does not add cross-domain coordination overhead, and Section~III-C describes the use of a compact \ac{SLM} to keep per-domain inference latency low.

From an intra-domain point of view, we place one agent at the top of each administrative resource domain, responsible for domain-local resource allocation in support of service provisioning. The domain agent continuously collects and summarizes local state (e.g., available compute/network capacity, current allocations, local policies, and observed \ac{QoS}), receives provisioning requests, and enforces actions such as placement, user--instance binding, scaling, and migration. It can also revise existing allocations to remain optimized as demand and infrastructure conditions change. The agent connects to the domain through an adapter layer: for modern, software-controlled resources it can interact directly via a unified tool interface (e.g., Model Context Protocol (MCP) connectors toward devices/controllers), while for legacy infrastructures it operates through existing management stacks such as \acp{NMS} or cloud/edge controllers that expose monitoring and actuation \acp{API}.

From an inter-domain point of view, the agents form an overlay control-plane \emph{graph}, where each node corresponds to a domain agent and each edge represents a communication relationship \cite{wittner_communication_2026}. We align this overlay with data-plane coupling: if two domains are connected (or interdependent) in the data plane, their agents are connected in the control plane and can exchange information about relevant cross-domain state. Importantly, by placing the coordination logic in the agents, domains are not required to implement new cross-domain coordination standards; instead, agents communicate using \ac{A2A} protocols on the control plane. As a result, the architecture directly improves: \textbf{scalability} by limiting coordination to goal-driven, neighbor-to-neighbor exchanges rather than raw global state collection; \textbf{simplicity} by shifting orchestration logic from fragile, manually engineered workflows to an agentic closed loop over tools and feedback; and \textbf{agility} by supporting plug-and-play join/leave at the domain level (a domain onboards by deploying its agent+adapters and establishing edges to adjacent domains).

\subsection{Decision Making}
\begin{figure}[t]
    \centering
    \includegraphics[width=0.85\linewidth]{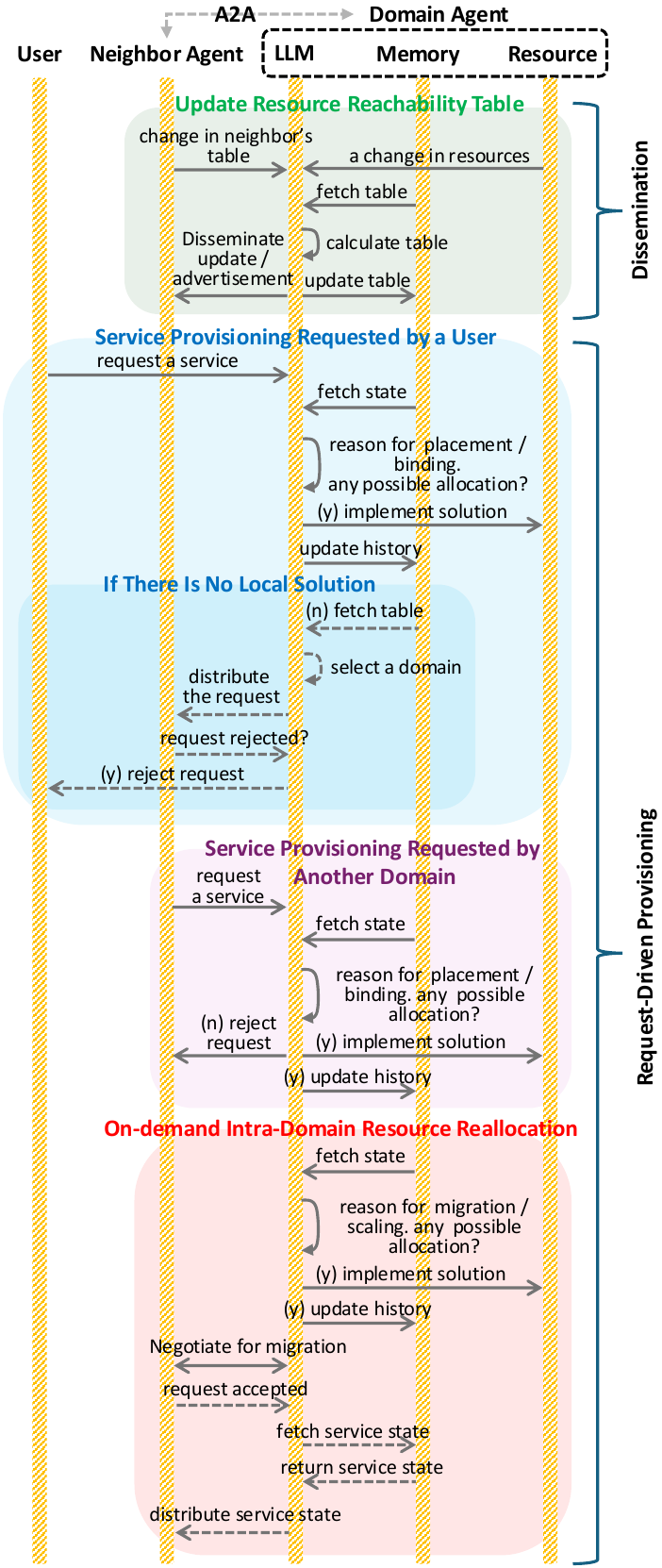}
    \vspace{-8pt}
    \caption{High-level procedures for change-triggered reachability dissemination and service provisioning on the inter-domain agent overlay. For tractability, the figure omits detailed per-hop mechanisms (e.g., multi-hop forwarding, soft reservations/commit).}
    \vspace{-0.5cm}
    \label{fig:decision_making}
\end{figure}

Our agents coordinate cross-domain provisioning using two complementary message-passing modes on the overlay control plane. The first is \emph{change-triggered dissemination}, which maintains a routing-like view of feasible resources; the second is \emph{on-demand request/negotiation}, which supports safe re-optimization and migration. In both cases, each domain agent remains authoritative only within its own administrative boundary: it advertises summarized reachability information, but enforces allocations (admission, placement, binding, scaling, migration) solely through its domain-local tool layer and policies. An overview is illustrated in Fig.~\ref{fig:decision_making}.

\textbf{Change-triggered dissemination (table-driven).} A domain agent summarizes the feasibility of reaching its \emph{local} computing resources from its data-plane \emph{inter-domain ingress/egress points}. Concretely, for each reachable compute resource it derives a compact \emph{resource advertisement} combining end-to-end latency, path bottleneck bandwidth, and available compute capacity. The agent then shares this advertisement with neighboring agents over \ac{A2A}. Upon receipt, a neighbor performs a routing-style update: for each advertised destination it adds its own intra-domain access latency to the destination, tightens the available bandwidth to reflect the new bottleneck, and records the next-hop neighbor toward that destination. This process repeats whenever a domain agent observes a material change in its intra-domain resources or connectivity (or receives an updated advertisement from a neighbor that changes its table). By repeating these local updates, agents build and refresh a distributed \emph{resource reachability table} that maps feasible remote computing resources to their predicted end-to-end latency and available capacities, without requiring raw global telemetry collection.

\textbf{Request-driven provisioning (per-request).} When a provisioning request arrives at a domain, the receiving agent first checks whether it can satisfy the request locally (subject to domain policies and current allocations). Otherwise, it consults its reachability table to select a feasible remote destination whose advertised compute and bandwidth capacities satisfy the request and whose predicted end-to-end latency meets the request’s \ac{QoS} constraints; among candidates, it prefers the one minimizing latency. The request is then forwarded hop-by-hop on the control plane until it reaches the destination domain agent. Each intermediate agent installs a \emph{soft reservation} on its segment and the outbound inter-domain link; once the destination domain allocates compute, a confirmation travels back and each domain \emph{commits} its reservation, binding the control-plane chain to an end-to-end data-plane path while preserving domain autonomy. Because feasibility information is already available in the table and reservations are established incrementally per hop, this mode supports low decision latency for per-request allocations while keeping coordination bounded to neighbor exchanges.

\textbf{On-demand negotiation (event-driven).} Disseminated tables are intentionally coarse: they enable fast, feasible placement, but they do not by themselves guarantee that subsequent \emph{changes} (e.g., intra-domain path rebalancing, resource reallocation, or migration) preserve each service’s \ac{QoS}. For these cases, agents switch to an on-demand mode in which an agent initiating a change (e.g., flow rerouting, migration) negotiates with the agent(s) responsible for the affected service instance(s) via \ac{A2A} messaging, exchanging the additional state needed (updated path characteristics, migration context) to confirm that end-to-end \ac{QoS} is preserved before enforcement. While such negotiation introduces additional control latency, it is used for optimization and lifecycle adaptation rather than the critical path of initial admission, thereby enabling improved optimality without imposing a centralized orchestrator.

\subsection{Training \& Inference}
To operate as domain-top controllers, the proposed agents must generate provisioning decisions that are \emph{fast} yet sufficiently \emph{intelligent} for multi-constraint optimization. Accordingly, each domain agent is instantiated with an \ac{SLM} \emph{pretrained on strong reasoning tasks} as its reasoning core (DeepSeek-R1-Distill-Qwen-7B in the evaluation): this choice preserves low inference latency while providing the reasoning competence needed for reliable, goal-driven control (training procedure and evaluation are detailed in what follows and in Section~IV).
However, service provisioning in Organic 6G requires more than generic reasoning: the model must internalize domain policies, optimize under multiple criteria (e.g., latency and resource efficiency), and produce \emph{actionable} allocations that satisfy \ac{QoS} constraints. To this end, the \ac{SLM} is specialized using \ac{RL} with \emph{verifiable} feedback signals, following the self-verification paradigm popularized by recent reasoning models \cite{deepseek_r1_2025}.

\begin{figure}[t]
    \centering
    \includegraphics[width=0.8\linewidth]{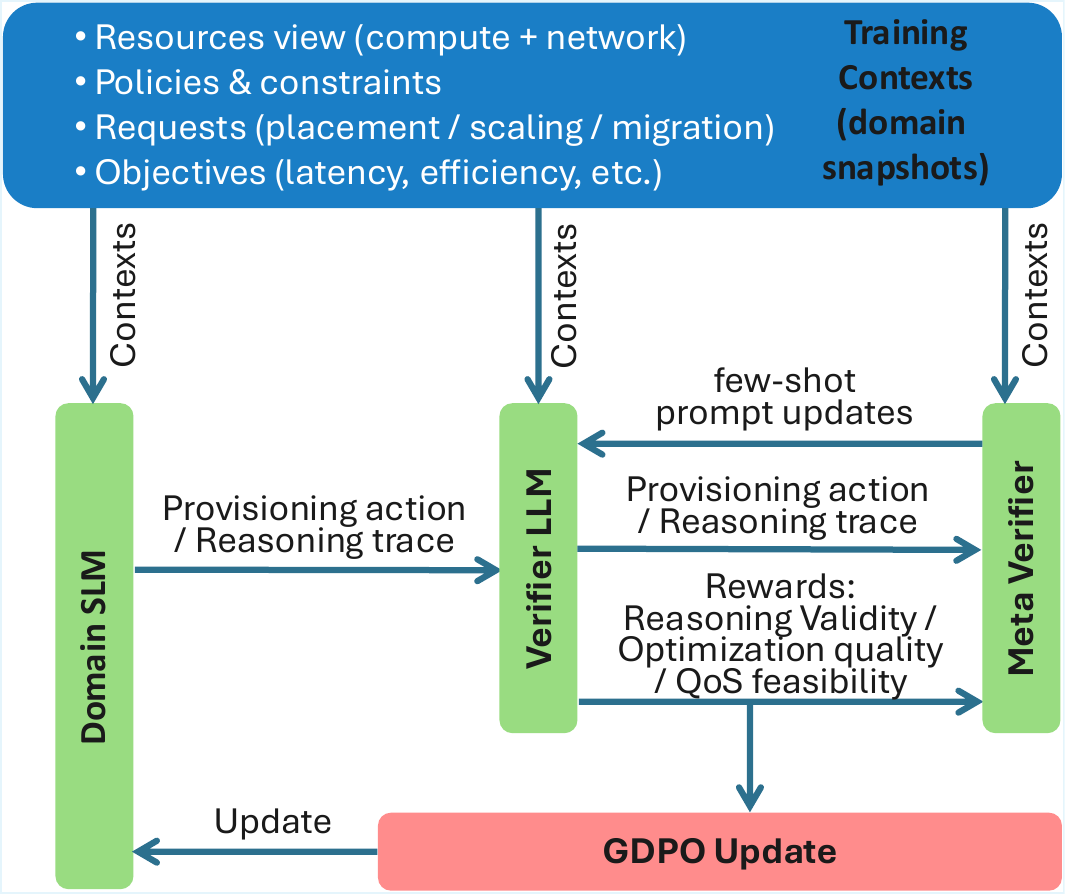}
    \vspace{-8pt}
    \caption{Offline self-verification training of domain \acp{SLM}. Training contexts (domain snapshots) elicit reasoning traces and provisioning actions; a strong verifier assigns multi-objective rewards (reasoning validity, optimization quality, and \ac{QoS} feasibility) that drive GDPO updates. Verifier reliability is improved via meta-verifier--guided few-shot prompt refinement.}
    \label{fig:offline_self_verification}
\end{figure}

\textbf{Offline self-verification training.} Fig.~\ref{fig:offline_self_verification} summarizes the offline self-verification procedure. A set of training \emph{contexts} is constructed to reflect the foundations in Section~II and the decision process in this section: each context encodes (i) a domain view of compute and network resources, (ii) local policies and constraints, (iii) representative provisioning requests, and (iv) target optimization criteria. For each context, the \ac{SLM} generates a reasoning trace and a candidate provisioning action (placement, user--instance binding, scaling, or migration). A stronger verifier \ac{LLM} then assesses the output and assigns a \emph{multi-objective} reward vector capturing three aspects: (1) \emph{reasoning validity} (is the allocation supported by a coherent and complete reasoning path, avoiding unjustified shortcuts \cite{xie_logic_rl_2025}), (2) \emph{optimization quality} (how well the proposed decision matches the declared objectives such as latency minimization under capacity constraints), and (3) \emph{\ac{QoS} feasibility} (whether the requests can be served within their \ac{QoS} bounds given the advertised resources). These heterogeneous rewards are used to update the \ac{SLM} with \emph{Group reward-Decoupled Normalization Policy Optimization (GDPO)}, which stabilizes training by normalizing each reward component before aggregation, thereby preserving learning signal resolution in multi-reward settings \cite{liu_gdpo_2026}. In parallel, the verifier is improved through a lightweight bootstrapping loop: a meta-verifier reviews verifier judgments, and the resulting feedback is injected into the verifier prompt (few-shot style) to progressively reduce systematic errors, without requiring additional training infrastructure.

\begin{figure}[t]
    \centering
    \includegraphics[width=0.85\linewidth]{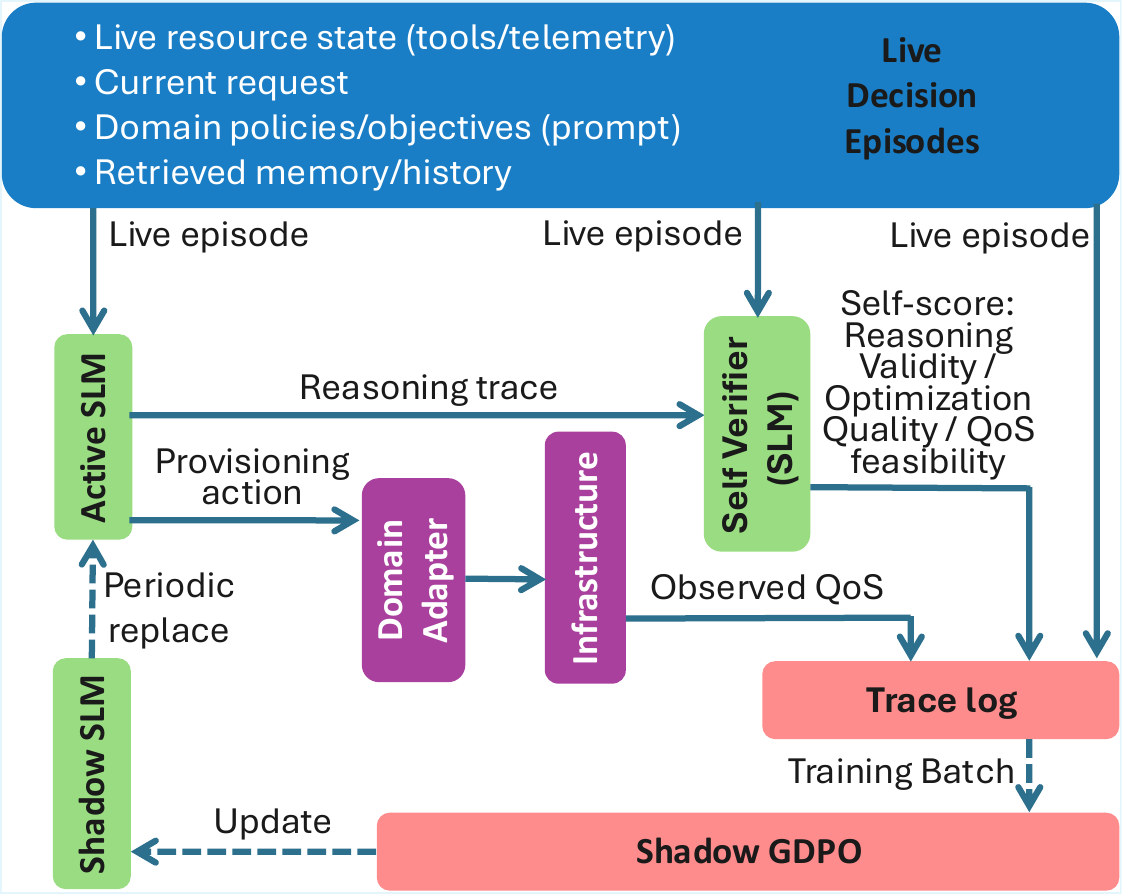}
    \vspace{-8pt}
    \caption{Online inference and periodic refinement with shadow updates. The deployed \ac{SLM} agent retrieves memory, queries live state via tools, enforces provisioning actions through the domain adapter, and logs traces together with observed \ac{QoS}. Traces are batched to update a shadow copy offline (e.g., GDPO), which periodically replaces the deployed model.}
    \vspace{-0.5cm}
    \label{fig:online_shadow_update}
\end{figure}

\textbf{Online inference and periodic refinement.} Fig.~\ref{fig:online_shadow_update} illustrates the online closed loop and the periodic shadow-update refinement strategy. After offline training, the \ac{SLM}-based agent is deployed on top of each domain with access to its memory and tool layer, and with a system prompt encoding the domain’s policies and optimization criteria. At inference time, the agent follows the closed loop described earlier: it retrieves relevant history, collects live state via tools, proposes a decision, and enforces it through the domain adapter. Reliability is improved by reusing the same self-verification loop as in offline training, which also serves as a lightweight hallucination guard: outputs violating \ac{QoS} feasibility or lacking coherent reasoning are penalized before enforcement. For each decision, the \ac{SLM} produces a reasoning trace and an allocation, then evaluates its own output using the same verifiable, multi-objective reward criteria learned during training. To adapt to evolving traffic and infrastructure conditions without blocking real-time control, inference traces (requests, state snapshots, actions, rewards, and observed \ac{QoS}) are accumulated into batches, and a \emph{separate} copy of the \ac{SLM} is updated periodically offline (e.g., with GDPO). From time to time, the updated copy replaces the working \ac{SLM} \cite{dang_rl_reasoning_small_llms_2025}. This \emph{shadow update} strategy decouples live decision making from continual retraining, while enabling steady improvement under non-stationary Organic 6G dynamics.

\section{Simulations}
\subsection{Scenario A: Control-Plane Evaluation}
\begin{figure}[!t]
    \centering
    \includegraphics[width=3.3in]{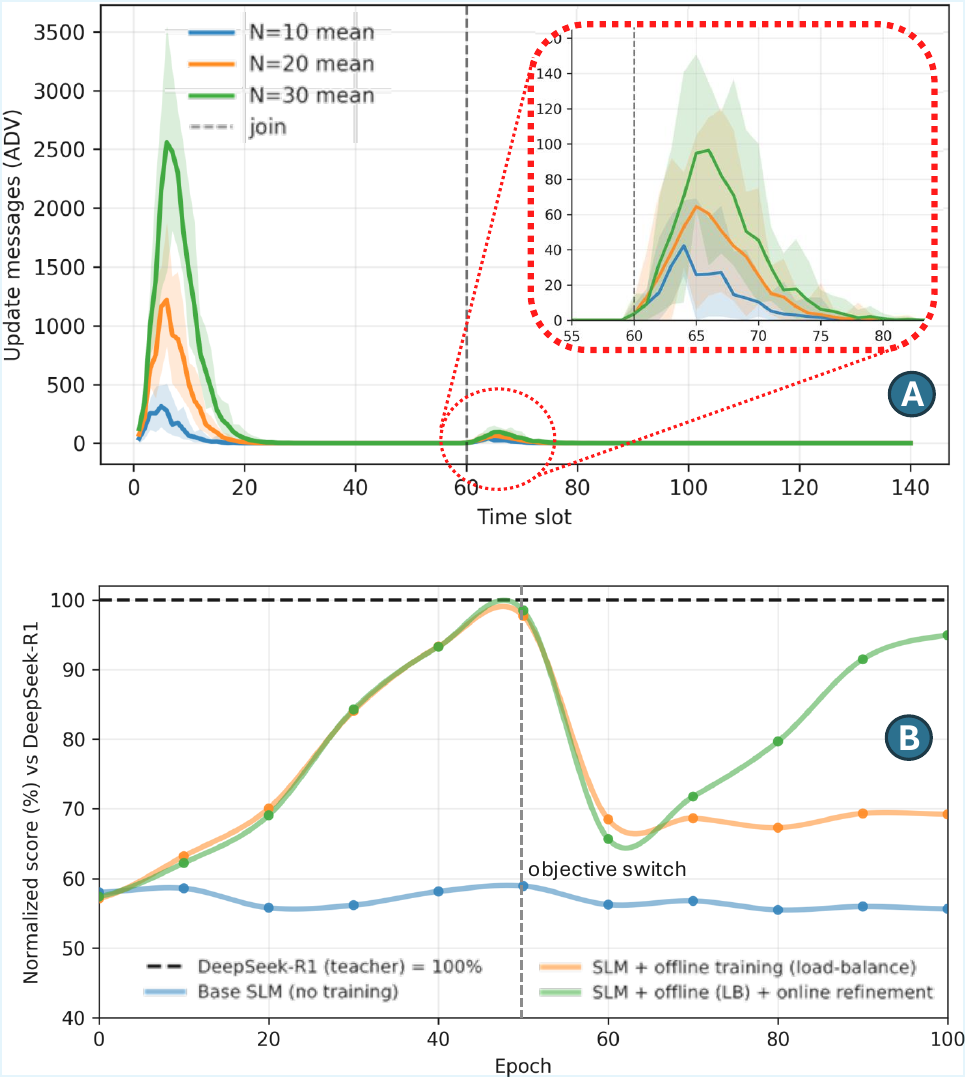}
    \vspace{-8pt}
    \caption{Simulation results for the two scenarios. (A) Scenario A: Message-passing overhead of distributed reachability-table dissemination, including a domain join at time slot 60. Lines show the mean number of update messages per time slot; shaded regions denote min--max across random topologies for each \(N\). (B) Scenario B: Normalized score (\%) relative to DeepSeek-R1 (verifier \ac{LLM} baseline fixed at 100\%). The \ac{SLM} improves with offline specialization on load-balance (epochs 0--50); after switching to a min-latency objective (epochs 50--100), only the variant with online refinement recovers toward the baseline.}
    \label{fig:fig6}
    \vspace{-12pt}
\end{figure}

We quantify \textit{control-plane message volume} to assess the overhead of the dissemination in Section~III, focusing on (i) reachability-table formation and (ii) re-convergence after a domain joins. The results are illustrated in Fig.~\ref{fig:fig6}.A. We simulate \(N\in\{10,20,30\}\) administrative domains whose agents form an overlay aligned with data-plane coupling. Each curve is averaged over 20 independent runs. In each run, \(N\) domains (each exposing 3 local compute resources summarized by its agent) are connected by a random inter-domain graph with target average degree 4, and the agents execute table-driven dissemination. Each time slot represents one management epoch (consistent with orchestration decision intervals on the order of seconds), so convergence within tens of slots corresponds to table formation on the order of minutes, appropriate for provisioning-level control. When a local table change occurs, the agent's message-preparation latency is measured and then normalized to a 1--3 time-slot interval before the update is emitted to neighbors; delivery succeeds with probability \(p=0.9\) and reliability is enforced by ACK-based retransmissions. To match the accounting in Fig.~\ref{fig:fig6}.A, each changed table entry is treated as a separate update message (sent in parallel within the same time slot), and we report update messages (ADV) only. The initial burst occurs while agents rapidly populate reachability tables with previously unknown destinations; once tables stabilize, change-triggered updates cease and the message rate collapses to near zero. At time slot 60, a new domain joins and attaches to a random subset of existing domains, introducing only incremental reachability entries, which yields a smaller transient before re-convergence. Overall, the overhead remains manageable, and for fixed average degree it increases approximately linearly with the number of domains, as more destinations must be disseminated over neighbor-only exchanges.

\subsection{Scenario B: Training and Refinement Evaluation}
This scenario validates the two key claims of Section~III-C: that offline \ac{RL} specialization brings the \ac{SLM} to near-verifier-level provisioning performance, and that online refinement enables recovery under objective changes. We evaluate the offline and online specialization loop using DeepSeek-R1 as the strong verifier \ac{LLM} and DeepSeek-R1-Distill-Qwen-7B as the deployed \ac{SLM} reasoning core, over a synthetic heterogeneous multi-domain infrastructure whose parameters (12 domains, inter-domain latencies 2--20\,ms, bandwidths 0.5--10\,Gbps, and per-domain compute pools spanning 32--128 vCPU-equivalents) span the plausible operating range of 6G deployments. We generate a dataset of 3000 service-provisioning scenarios over this infrastructure. Training proceeds in two phases. During epochs 0--50, the \ac{SLM} is specialized offline toward a \textit{load-balance} objective (balancing resource utilization). From epoch 50 onward, the objective switches to \textit{min-latency} (minimizing overall provisioning latency). Both phases are also subject to \ac{QoS} and capacity feasibility. We evaluate at checkpoints every 10 epochs and report a normalized score under the active objective, where DeepSeek-R1 is used as the 100\% reference. Fig.~\ref{fig:fig6}.B compares three variants: (i) a base \ac{SLM} without training, (ii) \ac{SLM} with offline training (trained for load-balance only), and (iii) \ac{SLM} with offline training followed by online refinement via periodic shadow updates. Offline training drives the \ac{SLM} close to the DeepSeek-R1 reference on the load-balance objective. After the objective switch, the offline-only variant drops and remains lower, whereas online refinement restores performance as the model adapts to the new min-latency criterion. Overall, the results indicate that the proposed offline+online specialization yields sufficient capability to maintain high accuracy relative to the verifier \ac{LLM} under both stationary operation and objective changes. Practically, achieving near-verifier-level scores on the min-latency objective means the deployed \ac{SLM} produces placement and binding decisions that approach the quality of a full \ac{LLM}, translating directly to tighter end-to-end service latency for users while keeping inference local to each domain. Note that Scenario A and Scenario B evaluate complementary aspects: the former assesses control-plane dissemination overhead (independent of the decision-maker), while the latter evaluates \ac{SLM} reasoning quality and adaptation.

\section{Conclusion \& Future Work}
In this paper, we studied service provisioning over an Organic 6G infrastructure by viewing the system as a heterogeneous edge--cloud--\ac{NTN} continuum composed of independently administered resource domains. We formulated the problem as a continuous placement-and-binding loop under stringent \ac{QoS} targets, subject to scalability, simplicity, and agility requirements. To address these, we proposed a decentralized architecture based on \ac{LLM}-based domain agents connected by an \ac{A2A} overlay graph: agents used table-driven dissemination to build reachability views for fast feasible placement, and switched to event-driven negotiation for safe re-optimization and migration. Simulations validated that neighbor-only dissemination induced manageable control-plane overhead, and that an \ac{SLM} reasoning core specialized via verifier-based self-verification and periodic refinement maintained high decision quality, including under objective changes.

Looking forward, we outline several fundamental research directions for making agentic orchestration a principled and robust foundation for Organic 6G:
\begin{itemize}
    \item \textbf{Formal cross-domain agentic orchestration:} develop formal models for cross-domain trade-offs and scalability limits (including domain-count thresholds at given performance targets); current simulations demonstrate feasibility up to 30 domains with near-linear overhead, but formal bounds remain open.
    \item \textbf{Uncertainty-aware decision theory:} design methods that represent uncertainty and provide tail-risk \ac{QoS} guarantees under non-stationary dynamics (e.g., fluctuating link latency or intermittent compute availability at near-edge nodes).
    \item \textbf{Security for agentic control planes:} establish threat models and defenses for \ac{A2A}-mediated orchestration (e.g., a malicious domain agent injecting false reachability advertisements to attract or divert traffic, confused-deputy behavior, or compromised tool responses), together with adversarial benchmarks.
    \item \textbf{Multi-agent stability and conflict resolution:} study agent-population conflicts and oscillations (e.g., two neighboring agents repeatedly migrating the same service instance back and forth, or a cascade of reallocations triggered by a single domain change), and design lightweight arbitration and convergence mechanisms.
    \item \textbf{Mechanism design for truthful coordination:} study incentive-compatible protocols for truthful resource advertisement (i.e., without incentive to overstate or understate available resources for strategic gain) and fair sharing, aligning local utilities with end-to-end \ac{QoS}.
    \item \textbf{Learning and adaptation under real-time constraints:} quantify adaptability--cost trade-offs: safe online learning, energy/latency budgets, and privacy-preserving or federated adaptation in heterogeneous environments.
\end{itemize}

\section*{Acknolwedgement}
This work was, in part, supported by BMFTR, Germany (6GEM+, Grant 16KIS2411), and the EU Horizon Europe programme (6G-Path, Grant 101139172).

\bibliographystyle{IEEEtran}
\bibliography{main}

\end{document}